 \documentclass[final,5p,times]{elsarticle}

\usepackage{amssymb}
 \usepackage{amsthm}

\usepackage{subcaption}
\usepackage{here}

\journal{Physics letter A}

\begin{document}
\begin{frontmatter}



\title{Optimization of quantum noise by completing the square of multiple interferometer outputs in quantum locking for gravitational wave detectors}


\author[label1]{Rika Yamada}
\author[label2]{Yutaro Enomoto}
\author[label3]{Atsushi Nishizawa}
\author[label4]{ Koji Nagano}
\author[label1]{ Sachiko Kuroyanagi}
\author[label4]{ Keiko Kokeyama}
\author[label5]{ Kentaro Komori}
\author[label2]{Yuta Michimura}
\author[label1]{ Takeo Naito}
\author[label1]{ Izumi Watanabe}
\author[label1]{ Taigen Morimoto}
\author[label2]{ Masaki Ando}
\author[label6]{ Akira Furusawa }
\author[label1]{ Seiji Kawamura}

\address[label1]{Department of Physics, Nagoya University, Furo-cho, Chikusa-ku, Nagoya, Aichi 464-8602, Japan}
\address[label2]{Department of Physics, University of Tokyo, Bunkyo, Tokyo 113-0033, Japan}
\address[label3]{Research Center for the Early Universe (RESCEU), School of Science, The University of Tokyo, Tokyo 113-0033, Japan}
\address[label4]{Institute for Cosmic Ray Research, The University of Tokyo, 5-1-5 Kashiwa-no-Ha, Kashiwa, Chiba 277-8582, Japan}
\address[label5]{LIGO Laboratory, Massachusetts Institute of Technology, Cambridge, Massachusetts 02139, USA}
\address[label6]{Department of Applied Physics, School of Engineering, The University of Tokyo, 7-3-1 Hongo, Bunkyo-ku, Tokyo 113-8656, Japan}

\begin{abstract}
The quantum locking technique, which uses additional short low-loss sub-cavities, is effective in reducing quantum noise in space gravitational wave antenna DECIGO. However, the quantum noise of the main interferometer depends on the control systems in the sub-cavities. Here we demonstrate a new method to optimize the quantum noise independently of the feedback gain by completing the square of multiple interferometer outputs in the quantum locking system. We successfully demonstrate in  simulations that this method is effective in optimizing the homodyne angle to the best quantum-noise-limited sensitivity.

\end{abstract}






\end{frontmatter}



\section{Introduction}
\label{intro}
Gravitational waves were detected for the first time by LIGO in 2015 \cite{GW150914}. Many detections of gravitational waves from binary black hole mergers and binary neutron star mergers have  followed since \cite{LIGO.o1o2.catalog}. Now, one of the next major targets is the detection of primordial gravitational waves, which are  believed to have been produced during the inflation period \cite{primordialGW}. Given that gravitational wave detection is the only  observable direct proof of the inflation, its significance cannot be overstated. The decihertz interferometer gravitational wave observatory (DECIGO) is  a Japanese future space mission,  of which the primary goal is to detect the primordial gravitational waves \cite{decigo1}. It  is a fleet of three  drag-free spacecraft 1,000 km apart from one another.

In interferometric gravitational wave detectors, quantum noise is one of the fundamental noise sources  that limit their sensitivity \cite{Braginsky1996}.Quantum noise consists of radiation pressure noise and shot noise. The sensitivity of DECIGO is limited by the radiation pressure noise at lower frequencies and by shot noise at higher frequency. The radiation pressure noise can be reduced by making the mirror heavier, but in case of DECIGO, the mirror mass is already 100 kg and it is difficult to make the mirror drastically heavier because of the weight limit of the payload. To reduce quantum noise, squeezed light is used in ground-based detectors \cite{GEO300, LIGO, Virgo}.	However, the arms of DECIGO are 1,000 km long and optical loss due to diffraction would be too large (22 \% for each mirror reflection) to use squeezed light.  Instead, the quantum locking technique \cite{qlock1,qlock2} could be employed in DECIGO,  which is an effective technique in reducing the quantum noise of a cavity with high optical loss, utilizing short low-loss cavities. Incidentally, the quantum locking technique has not been experimentally demonstrated so far.

To use quantum locking, short sub-cavities are required to be implemented on the outer sides of the main cavity's mirrors. Locking of the sub-cavity length to the laser by acting on the main mirror yields reduction of the radiation pressure noise of the main mirrors,  providing that the sub-cavities are operated with a lower laser power than the main cavities.  In addition, the radiation pressure noise of the main mirrors can be also eliminated at a certain frequency.  To achieve this, homodyne detection must be implemented with a proper homodyne angle in the sub-cavities.  Also,  the homodyne angle of homodyne detection should be optimized for  the particular frequency determined  on the basis of each science target,  which is possible only with  precise estimation of the quantum noise  as a function of the homodyne angle. The main problem is, however, that the quantum noise of the main interferometer depends on the control systems in the sub-cavities \cite{Furusawanature} because the control systems impose the sub-cavity's quantum noise to the main cavity's mirrors. Therefore, it is  vital to develop a method to estimate the quantum noise independently of the control systems.

In this paper, we show a new method for this in which the best combination of  outputs from main- and sub-cavities is taken into account. We then demonstrate that the best homodyne angle for the quantum noise can be determined with this method.

\section{Theory}
\label{theo}
\begin{figure}
\includegraphics[width=120mm,bb=  0 0 926 441]{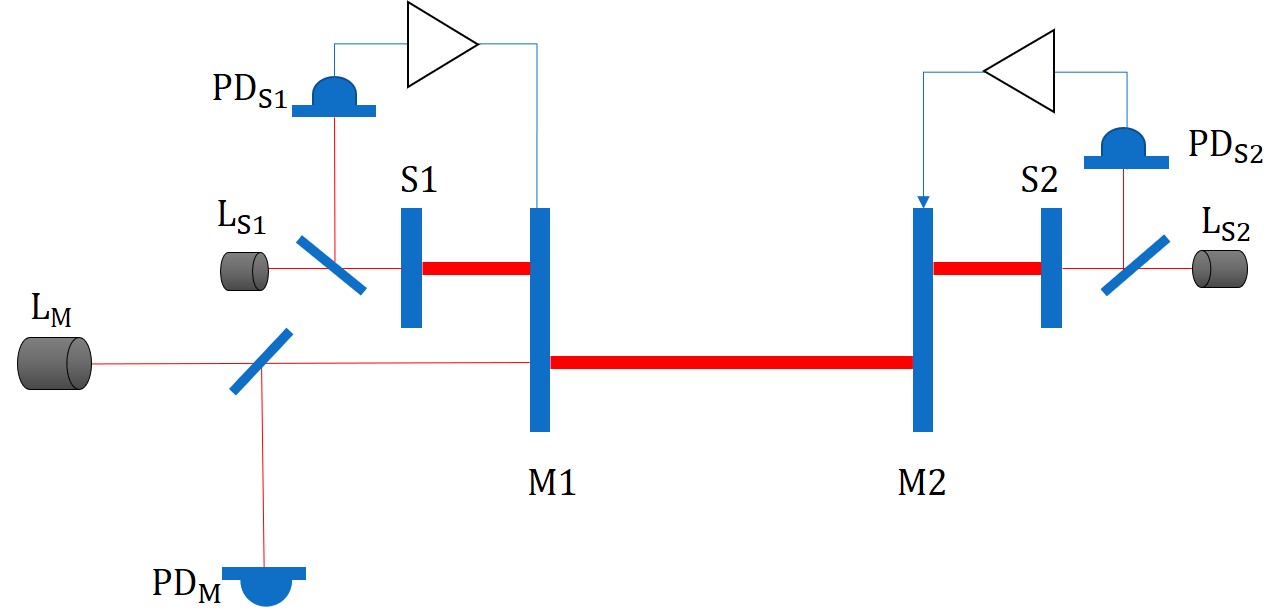}
\caption{Basic configuration of quantum locking. A main cavity consisting of two mirrors (M1, M2) is illuminated by a laser (${\rm L_M}$) and the reflected light is detected by a photodetector (${\rm PD_{M}}$). Two sub-cavities (sub-cavity 1, sub-cavity 2) consisting of the shared mirrors (M1, M2) and additional mirrors (S1, S2) are illuminated by two lasers (${\rm L_{S1}}$, ${\rm L_{S2}}$) and the reflected beams are detected by the two respective photodetectors (${\rm PD_{S1}, PD_{S2}}$). }
\label{cavity}
\end{figure}

Figure \ref{cavity} illustrates the basic configuration of quantum locking; the sub-cavities are implemented on the outer sides of the main cavity'{}s mirrors. The two mirrors of the main cavity are shared by each sub-cavity. Each sub-cavity (sub-cavity 1, sub-cavity 2) is locked on resonance by controlling the corresponding main mirror.

\begin{figure}[htb]
\includegraphics[width=80mm,bb=0 0 423 455]{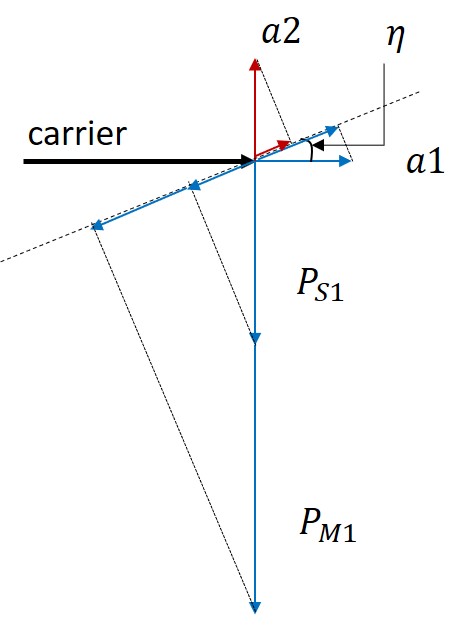}
\caption{Diagram at the detection port of  a sub-cavity.  ``Carrier'' indicates the laser carrier of the sub-cavity. $q_1$ is the amplitude quadrature and $p_1$ is the phase quadrature. $P_{\rm S1}$ and $P_{\rm M1}$ are the phase fluctuations of the reflected light  caused by the motion of the mirrors,  which are shaken by the amplitude quadrature.  $\eta$ is  the homodyne angle of the homodyne detection. Projected signals along the homodyne axis are also shown.}
\label{phaser}
\end{figure}

Figure \ref{phaser} shows a phasor diagram at the sub-cavity detection port.We use quadrature-phase amplitudes to describe quantum fluctuation throughout the paper \cite{two-phot}. Here let us consider field operators $a$ and $a^\dagger$, $[a, a^\dagger] = 1$, which are annihilation and creation operators. With $\hbar=\frac{1}{2}$ (photon number units), we define hermitian operators, q and p as $a=q+ip$ $q$ is called the amplitude quadrature and $p$ is called the phase quadrature. Note that $q_0$ and $p_0$ are for the main cavity, $q_1$ and $p_1$ are for the sub-cavity 1, and $q_2$ and $p_2$ are for the sub-cavity 2.
When  laser light enters  a sub-cavity,  the phase fluctuation of the light is  further affected by the amplitude and phase quadrature, respectively.

 When the laser light hits one of the mirrors in the sub-cavity, the mirror is physically shaken by the amplitude quadrature, $q_1$, coupling with the carrier amplitude. As a result, phase fluctuation of the reflected light, so-called radiation pressure noise ($ P_{\rm M1}$  and $ P_{\rm S1}$ for  mirrors M1 and S1, respectively), is introduced. 

Homodyne detection is a method to detect different quadrature components along a specific axis determined by the homodyne angle. We use local light, which has the same frequency as the carrier light but a different phase in the homodyne detection. We can rotate the detection axis by adding the local light to the carrier; we can detect signals projected on the dashed line with homodyne angle of $\eta$ (Fig.\ref{phaser}). With this  method, we can  adjust the parameters so that the phase fluctuation of the S1 mirror ($ P_{\rm S1}$) and the amplitude quadrature $q_1$ cancel each other at the detection output at a certain frequency.   
 This means that we detect  the phase fluctuation of the M1 mirror ($ P_{\rm M1}$) only. If we feed the detected signals of the sub-cavities back to the main cavity mirrors, we can completely eliminate the radiation pressure noise  in the main cavity at a certain frequency for a given homodyne angle.

 At high frequencies, however, the main cavity's quantum noise increases because the control system adds the sub-cavity's noise caused by $q_1$ to the main cavity's mirror. At low frequencies, the quantum noise depends on   to what extent the radiation pressure noise of the main cavity is replaced by that of the sub-cavity.  As such, the quantum noise of the main cavity depends on the control system.

 The best estimate of the quantum noise that is independent of the control system can be made by taking the best combination of the main cavity's output and the sub-cavities' outputs. Here, we obtain the best combination by completing the square.

\begin{figure}[htb]
\includegraphics[width=100mm,bb= 0 0 701 527]{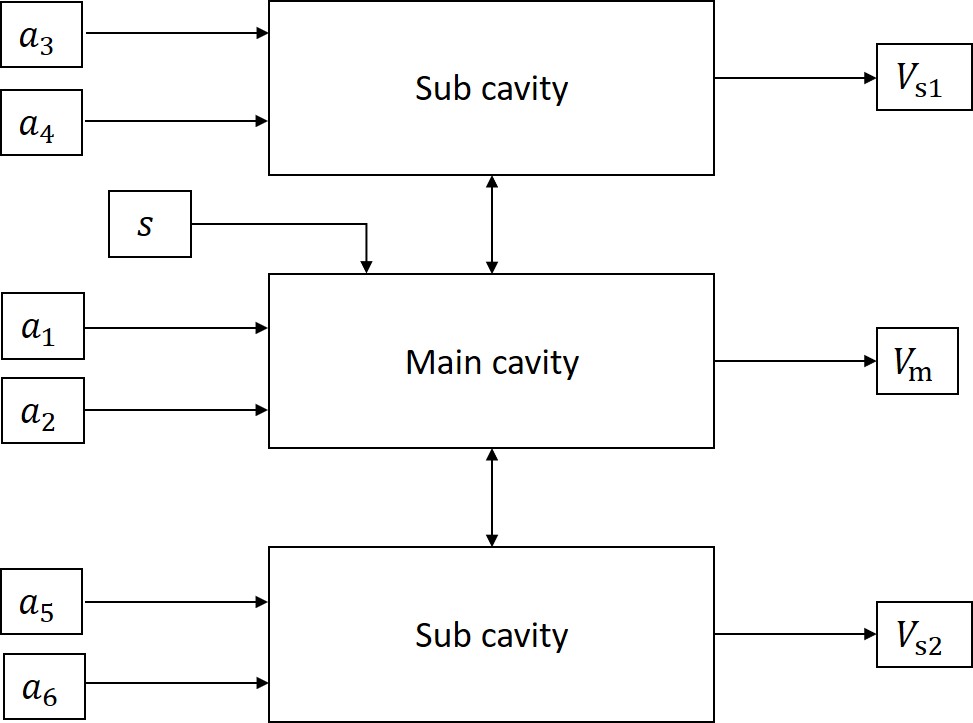}
\caption{Schematic block diagram of quantum locking. There are three cavities:  a main cavity and  two sub-cavities. 
 $q_0, q_1$, and $q_2$ are the amplitude quantum fluctuations. $p_0, p_1$, and $p_2$ are the phase quantum fluctuations.
 $V_{\rm m}, V_{\rm s1}$, and $V_{\rm s2}$ are the signals at the detection ports. S is  the  input (gravitational wave) signal.}
\label{block1}
\end{figure}

Figure \ref{block1} is a schematic block diagram, which shows three cavities of the quantum locking. Each cavity has one detection port ($V_{\rm m}, V_{\rm s1}, V_{\rm s2}$), and two quantum noise inputs.
Note that the main and sub-cavities are assumed to have end mirrors with 100 \% reflectivity; we neglect the vacuum fluctuations entering in place of the cavity losses. This assumption is used to demonstrate the effectiveness of the optimization method without any complications.
Gravitational wave signals enter only the main cavity as the sub-cavities are too small to be affected by gravitational waves.
 The outputs  are  given by Eqs.(\ref{Vm})-(\ref{Vs2}):

\begin{eqnarray}
Vm &=& s + A q_0 + i B p_0 + C q_1 + i D p_1 + E q_2 + i F p_2 \label{Vm} \\
Vs1 &=& G q_0 + i H p_0 + I q_1 + i J p_1 \\
Vs2 &=& K q_0 + i L p_0 + M q_2 + i N p_2. 
\label{Vs2}
\end{eqnarray}

The parameters $A$ to $N$ are the coefficients of each noise source. Note that A to F are normalized so that the coefficient of  $s$ is unity in $V_{\rm m}$. The parameters $V_{\rm s1}$ and $V_{\rm s2}$ are symmetric: $G=K$, $H=L$, $I=M$ and $J=N$.
 Let us introduce the combined output $V$ with a coefficient $\chi$,   as defined in Eq.(\ref{V}):

\begin{equation}
V = V_m + \chi \left( V_{s1} + V_{s2} \right).
\label{V}
\end{equation}

Here let us consider a quadrature sum of $V$, $x^2$, to estimate the signal to noise ratio. First, the squares of $q_0$, $q_1$, $q_2$ and $p_0$, $p_1$, $p_2$ are all $1/4$. Secondly, since all modes are bosonic modes, we can neglect all the cross-terms between them. Therefore, the quadrature sum, $x^2$, is derived as follows (Eq.(\ref{total})):

\tiny
\begin{eqnarray}
x^2 = s^2 + \{ \left( |2G|^2 + |2H|^2 +2 |I|^2 +2 |J|^2 \right) \left( \chi + \frac{2(AG^* + BH^* + CI^* + DJ^*)}{ \left( |2G|^2 + |2H|^2 +2 |I|^2 + 2|J|^2 \right)} \right) \left( \chi^* + \frac{2(A^*G + B^*H + C^*I + D^*J)}{ \left( |2G|^2 + |2H|^2 + 2|I|^2 + 2|J|^2 \right)} \right) \} \nonumber \\ + \{ - \frac{4 | AG^* + BH^* + CI^* + DJ^* |^2}{ \left( |2G|^2 + |2H|^2 +2 |I|^2 + 2|J|^2 \right)}+ \left( |A|^2 + |B|^2 + |C|^2 + |D|^2 + |E|^2 + |F|^2 \right) \}.
\label{total}
\end{eqnarray}
\normalsize

We have rearranged the terms of $x^2$ in order to complete the square with respect to $\chi$. Here, if we choose $\chi$ in such a way that the second term of $x^2$ is 0, $x^2$ takes the smallest value and accordingly the third term is the minimum value of $x^2$.
Therefore, the square root of the third term is the minimum total noise normalized in terms of  gravitational wave signals. 
%
%
\section{Simulations}

\subsection{Simulation model}
\label{Sim1}

 Following the previous section, where we have optimized the total noise of the main and sub-cavities using the completing-square method,  we  here determine the best homodyne angle for our  target frequency.

\begin{figure}[H]
\includegraphics[width=110mm,bb= 0 0 713 793]{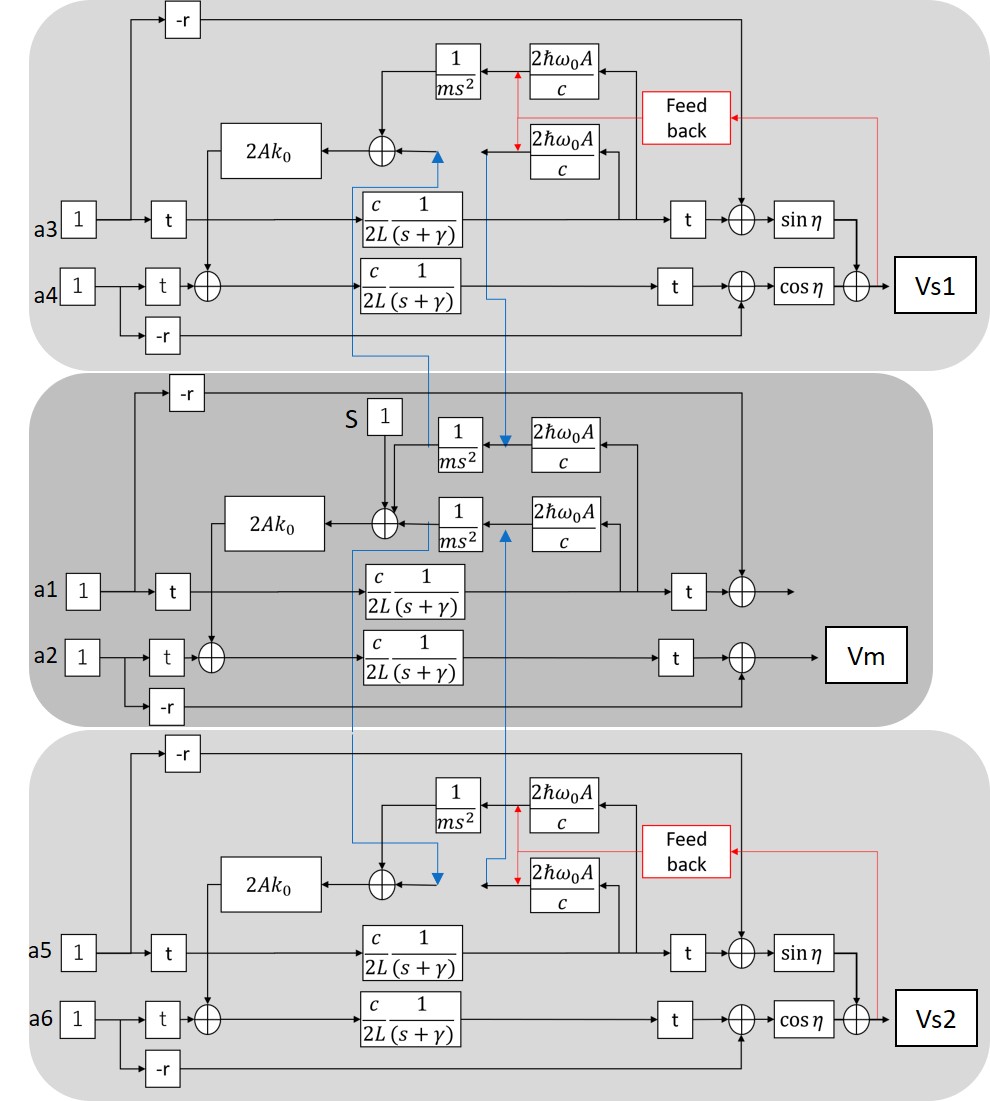}
\caption{Detailed block diagram of the main cavity and the two sub-cavities. The amplitude quadrature $q_0$ is divided into transmission and reflection by t and r blocks.  In the $c/2L(s+\gamma)$ block, the transmission is  affected by the cavity pole inside the cavity. The $2 \hbar \omega_0 A_0 /c$ block combines the amplitude quantum fluctuation with the carrier light, producing force applied to the mirrors. The $1/ms^2$ block  changes the force to the mirror displacement, which then produces  a phase fluctuation by $2A_0 k_0$. It is added to the phase quantum fluctuation. The phase quantum fluctuation is  affected by the cavity pole again.
Homodyne detection is  performed  in the $\sin{\eta}$ and $\cos{\eta}$ blocks. Finally it is detected at Vm.  Since the mirrors of the main cavity are shared with the sub-cavities, the $1/ms^2$ blocks are also shared. The feedback signal is added as a force which acts on the mirrors of the sub-cavities. }
\label{block2}
\end{figure}

Figure.\ref{block2} shows a block diagram (similar to but more detailed than Fig.~\ref{block1}) that demonstrates a procedure to obtain the coefficients $A$ to $J$ in Eq.(\ref{Vm})-(\ref{Vs2})  (see also \cite{EnomotoMron}). The center part is the main cavity. The amplitude quadrature ($q_0$) and the phase quadrature ($p_0$) are divided into the transmission and reflection by the input mirror of the main cavity with  an amplitude transmissivity (t) and amplitude reflectivity (r) of the mirror. The mirror  is assumed to have no optical loss: $t^2+r^2=1$.  The fluctuations are low-pass-filtered inside the cavity with a transfer function of the form of $c/2L(s+\gamma)$, where  $s$ is the Laplace complex variable and $\gamma$ is  the cavity pole. Note that we consider the case where each cavity is over-coupled.
Specifically, $\gamma$ is given by

\begin{eqnarray}
\gamma &=& \frac{\pi c}{2 L {\mathcal F}} \\
{\mathcal F} &=& \frac{\pi \sqrt{r}}{1-r},
\end{eqnarray}
 where $c$ is the speed of light, $L$ is the arm length of the cavity, and ${\mathcal F}$ is the  finesse of the cavity. 
The amplitude quadrature is coupled with the carrier light  with $2 \hbar \omega_0 A_0 /c$, where $\hbar$ is the reduced Planck constant, $\omega_0$ is the angular frequency of the light $\omega_0 = \frac{2 \pi c}{\lambda}$, and $A_0$ is the amplitude of light $A_0 = \sqrt{\frac{2 I_{\rm in}}{\omega_0 \hbar}}$,  which is a function of  the intensity ($I_{\rm in}$) of the light.

This force shakes the cavity mirrors with $1/ms^2$, where m is the mass of the mirror. The mirror motion causes  a phase fluctuations with $2A_0 k_0$, where $k_0$ is the wavenumber ($k_0=\omega_0/c$). 

In the main cavity, a phase quadrature of the reflected light, Vm, is detected. The feedback block of the main cavity is ignored for simplicity. The gravitational wave signal enters the block as a differential displacement of the mirrors. 

The sub-cavities have basically the same block diagram as the main cavity. In quantum locking, the mirrors of the main cavity are shared by the sub-cavities. We detect the output signals ($V_{\rm s1}$ and $V_{\rm s2}$) of the sub-cavities, and feed them back to the mirror (M1 and M2 in Fig.\ref{cavity}) motion. Homodyne detection is  performed  with $\sin{\eta}$ and $\cos{\eta}$, where $\eta$ is the homodyne angle.

In  our simulations, we use optical and mechanical parameters  on the basis of the pre-conceptual design parameters of DECIGO  listed in table 1. The sub cavities' parameters are shown in table 2.

\begin{table}[htb]
\begin{center}
\caption{Pre-conceptual design parameters of the main cavity of DECIGO}
\begin{tabular}{|c|c|c|}\hline
 Cavity length& $L$ & 1000 km \\ \hline
 Finesse& ${\mathcal F}$ & 10 \\ \hline
 Laser Power& $P$ & 100 W \\ \hline
 Laser wavelength& $\lambda$ & 515 nm \\ \hline
 Mirror mass& $M$ & 100 kg  \\ \hline
\end{tabular}
\end{center}
\label{T1}
\end{table}


\begin{table}[htb]
\begin{center}
\caption{Parameters of the sub-cavities }
\begin{tabular}{|c|c|c|}\hline
 Cavity length& $L$ & 1 m \\ \hline
 Finesse& ${\mathcal F}$ & 10 \\ \hline
 Laser Power& $P$ & 100 W \\ \hline
 Laser wavelength& $\lambda$ & 515 nm \\ \hline
 Mirror mass& $M$ & 100 kg  \\ \hline
\end{tabular}
\end{center}
\label{sub}
\end{table}

Although the laser power is 10 W in the pre-conceptual design of DECIGO, 100 W is used in  our simulations because  shot noise is reduced with the latter  and hence it enables us to demonstrate the effect of the quantum locking more clearly.

\subsection{Optimization of quantum noise}
\label{Sim2}

\begin{figure}[hp]
\begin{tabular}{c}
\begin{minipage}{1 \hsize}
\includegraphics[width=80mm, bb=  0 0 420 315]{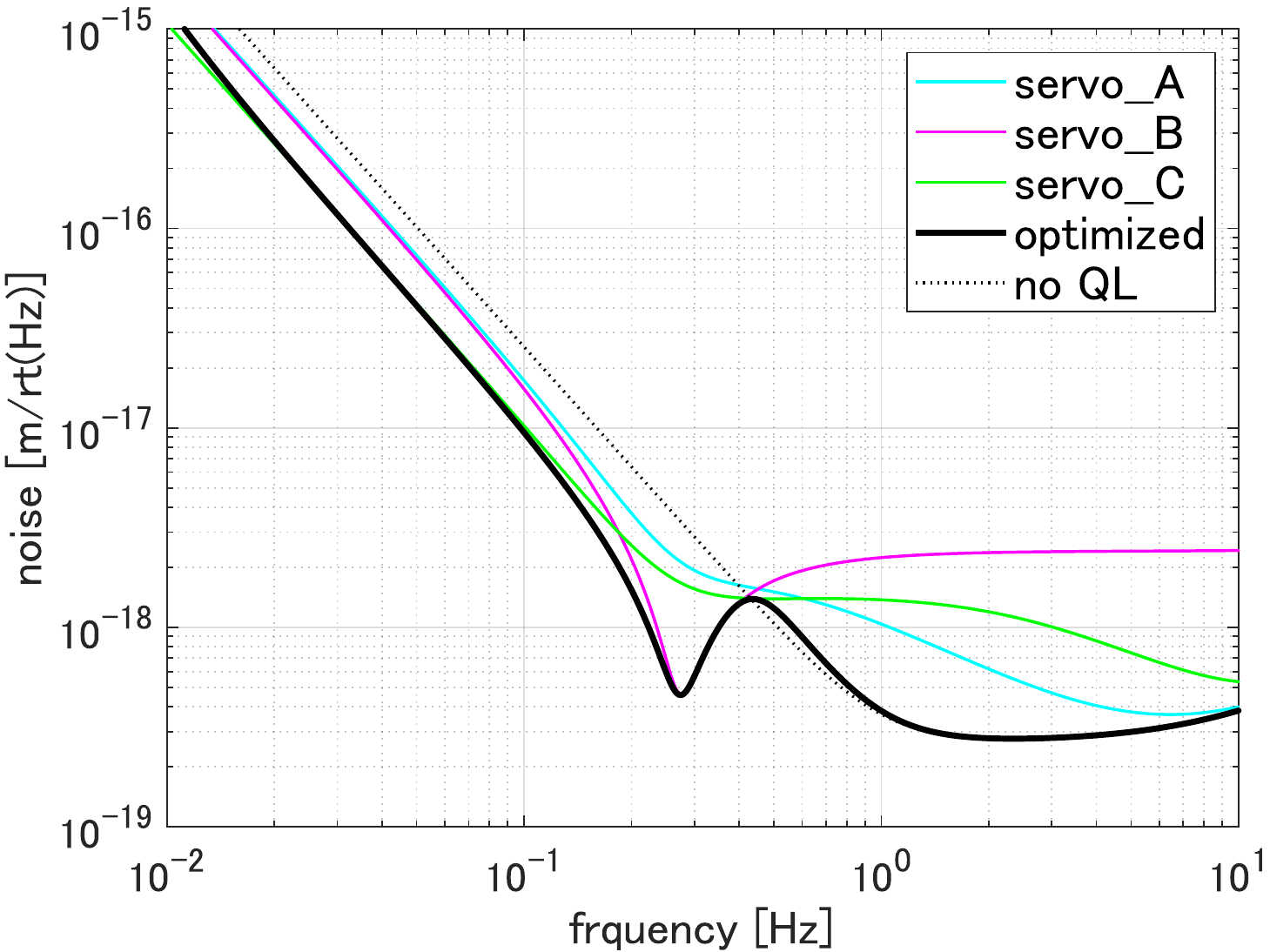}
\subcaption{}
\label{noise_curves}
\end{minipage}\\
\begin{minipage}{1 \hsize}
\includegraphics[width=80mm, bb=0 0 420 315]{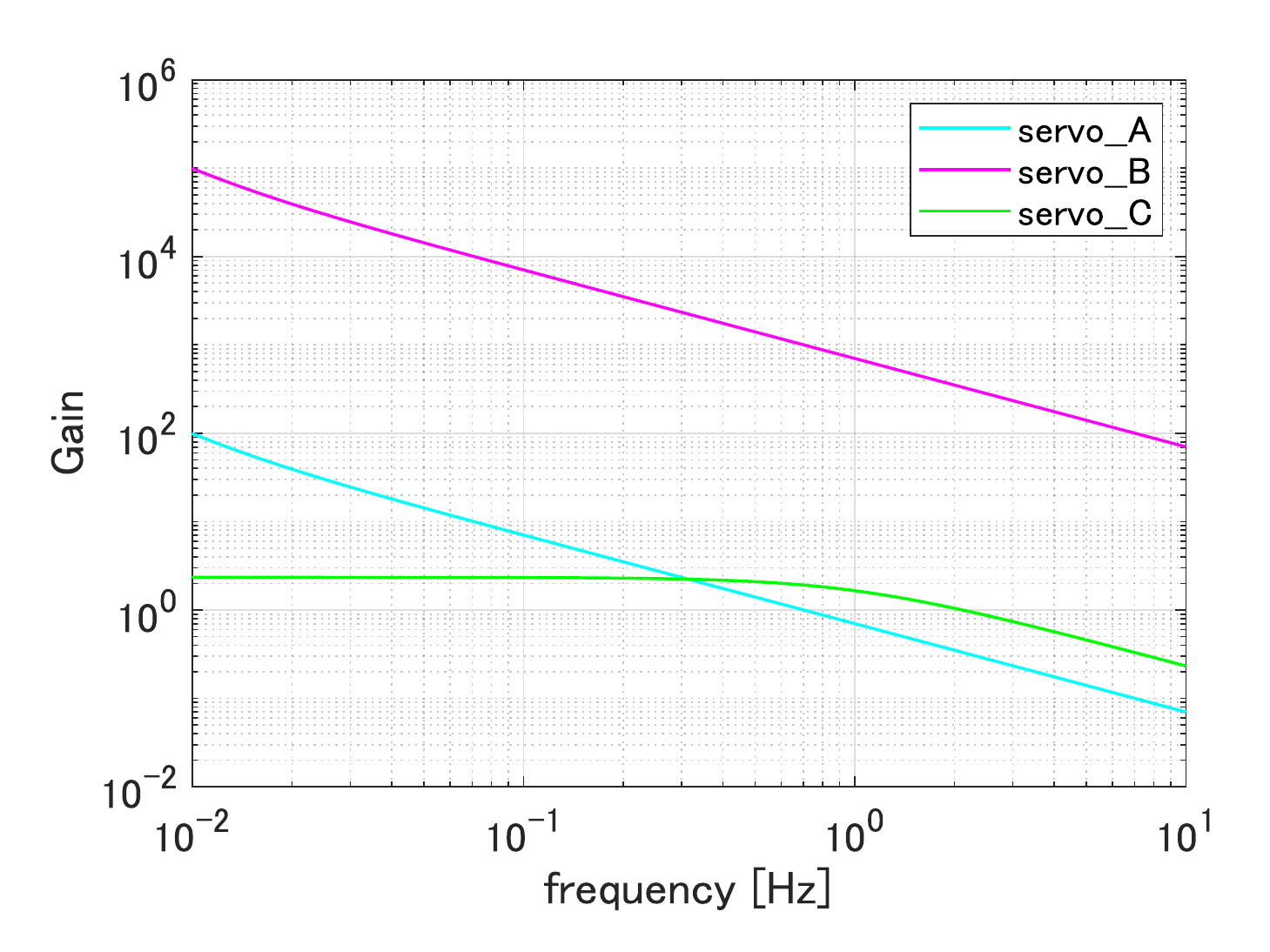}
\subcaption{}
\label{TF}
\end{minipage}
\end{tabular}
\caption{(a) Total quantum noise of the system. Bold black line shows the total noise, which is optimized  with the completing-square method. Colored solid lines show the quantum noise at the output of the main cavity with three variations of the servo system: servos A, B, and C. Dotted line (``no QL'') shows the total noise without the quantum locking. (b) Loop gains of the three servo systems of the sub-cavities as examples. }
\end{figure}

Using  our optimizing method (section \ref{theo}) and  model (section \ref{Sim1}), we optimize the quantum noise in the system with quantum locking. We calculate the noise with several feedback-loop gains (Fig.\ref{TF}) and show it in Fig.\ref{noise_curves}. Servo A has a moderate gain at low frequencies, servo B has a high gain at high frequencies and servo C has a gain slightly  higher than the unity at low frequencies. 
 If we use the output of only the main cavity, the total quantum noise depends on the feedback gain  (see servos A to C in Fig.\ref{noise_curves}). With servo A, the noise is lower at high frequencies, but the noise reduction at the most sensitive part is limited. With servo B, we  achieve the  best noise reduction at $\sim$0.3~Hz. However, the noise is higher for most of the other frequencies. With servo C, the noise is  lowest among the three servos at low frequencies.  
By contrast, the optimized total quantum noise  with the completing-square method shows remarkable overall improvement  (Fig.\ref{noise_curves});    it is as low as  that given by servo A at high frequencies, as low as  that by servo B at the  middle and best frequency ($\sim$0.3~Hz), and as low as  that by servo C at low frequencies. The optimized noise is the same even if the feedback of the sub cavity is turned off. This is the demonstration that the optimization is independent of any feedback gain used.

\begin{figure}[htb]
\includegraphics[width=80mm, bb=  0 0 420 315]{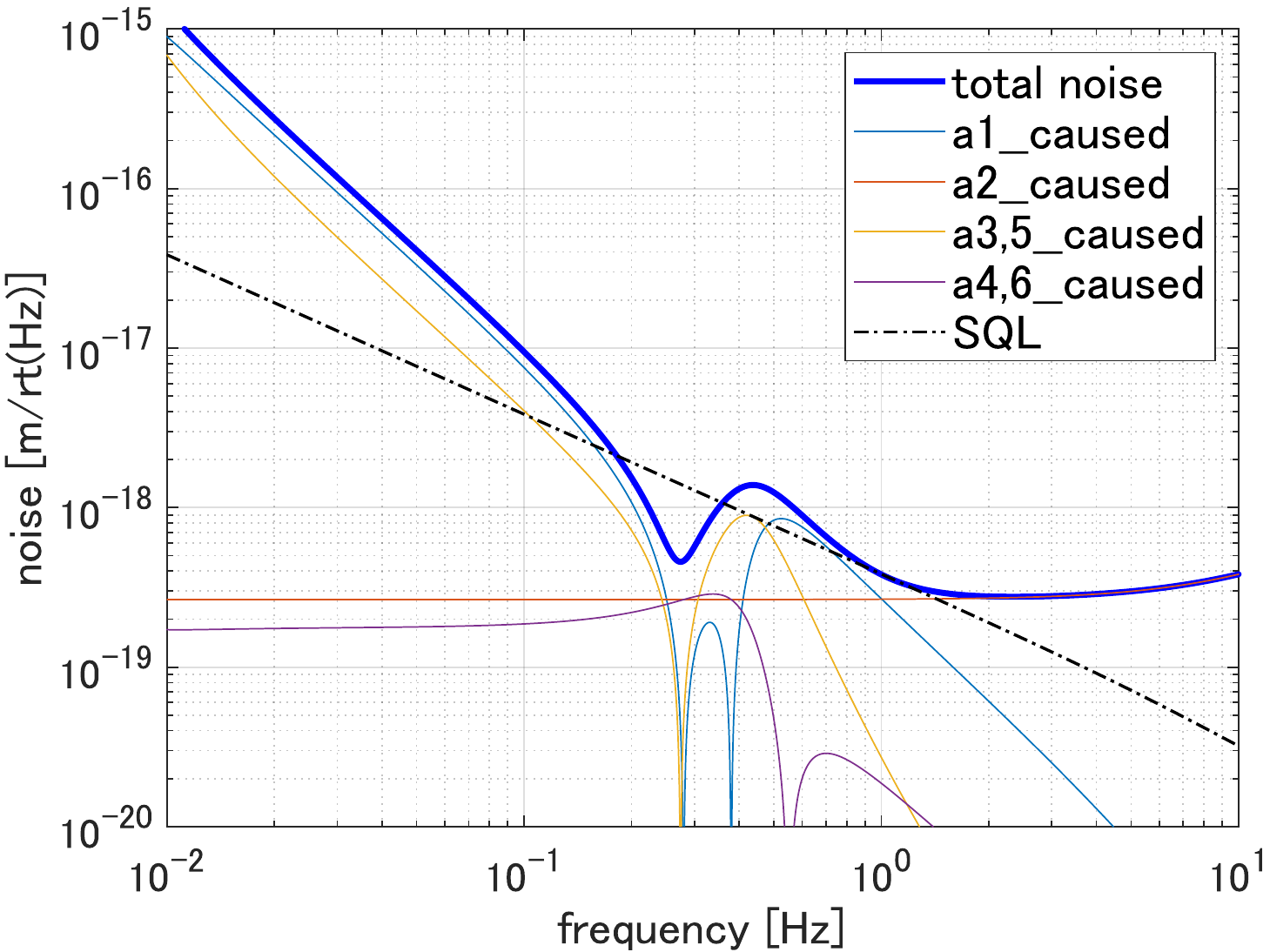}
\caption{Noise components of the optimized total quantum noise. The total noise is the sum of $q_{\rm 0\_caused}$, $q_{\rm 1\_caused}$, $q_{\rm 2\_caused}$ and $p_{\rm 0\_caused}$,  $p_{\rm 0\_caused}$,  $p_{\rm 1\_caused}$,  $p_{\rm 2\_caused}$ noises.  ``SQL'' is the standard quantum limit.}
\label{noise_detail}
\end{figure}

Figure \ref{noise_detail} shows the noise budget of the optimized total noise. Noise $p_{\rm 0\_caused}$  limits the total noise at high frequencies only. At the  middle and best frequency, noises $q_{\rm 0\_caused}$, $q_{\rm 1\_caused}$ and $q_{\rm 2\_caused}$   show  significant improvement,  whereas the improvement in noises  $p_{\rm 0\_caused}$, $p_{\rm 1\_caused}$ and $p_{\rm 2\_caused}$  is limited. At low frequencies,  noises $q_{\rm 0\_caused}$ and $p_{\rm 1,2\_caused}$  limit the total noise. This helps us understand which noise limits the sensitivity. This is useful in considering a strategy to improve the sensitivity.

\subsection{Signal to noise ratio}
We look for the homodyne angle which optimizes the total noise by maximizing the SNR with respect to primordial gravitational waves, our target source. We evaluate the total noise level  in the form of the signal-to-noise ratio (SNR) \cite{Braginsky1996} for primordial gravitational waves, according to  Eq(\ref{snr_eq}):

\begin{equation}
SNR = \frac{3 H_0^2}{10 \pi^2} \sqrt{T} \left[\int_{0.1}^{1} df  \frac{2 \gamma(f)^2 \Omega_{GW}^2(f)}{f^6 P_1(f) P_2(f)} \right]^{1/2}.
\label{snr_eq}
\end{equation}

\begin{table}[htb]
\begin{center}
\caption{Parameters  used to estimate the SNR}
\begin{tabular}{|c|c|c|}\hline
$H_0$ & Hubble parameter &  70 ${\rm km \cdot sec^{-1} \cdot Mpc^{-1}} $  \\ \hline
$T$ & time for correlation & 3 years \\ \hline
$f$ & frequency & 0.1 to 1 Hz \\ \hline
$\gamma$ & correlation function &1  \\ \hline
$\Omega_{GW}$  & energy density & $10^{-16}$  \\ \hline
$P_1, P_2$ & noise  power spectral densities & optimized in Figure 5 \\ \hline 
\end{tabular}
\end{center}
\label{table3}
\end{table}

$P_1$ and $P_2$ are the total noises, which are calculated in section \ref{Sim2}. Here we assume  them to be equal to  each other.  $T$ is the observation time, and is set to 3 years, as DECIGO plans to take a 3-year correlation. $\Omega_{GW}$ is the energy density upper limit of the primordial gravitational waves in the standard inflation model \cite{Planck2018, Kuroyanagi2014}. The target frequency band is chosen to be 0.1 to 1 Hz because confusion limiting noise, caused by unresolvable gravitational waves coming from binaries of white dwarfs etc., is significant below 0.1 Hz.  This means that we cannot detect the primordial gravitational waves at  frequencies lower than 0.1~Hz. Table 3 summarizes these parameters. In consequence, the SNR  at the output of only the main cavity without the sub-cavities is calculated to be 1.41.

\begin{figure}
\begin{tabular}{cc}
 \begin{minipage}{1 \hsize}
 \centering
 \includegraphics[width=50mm, bb= 0 0 420 315]{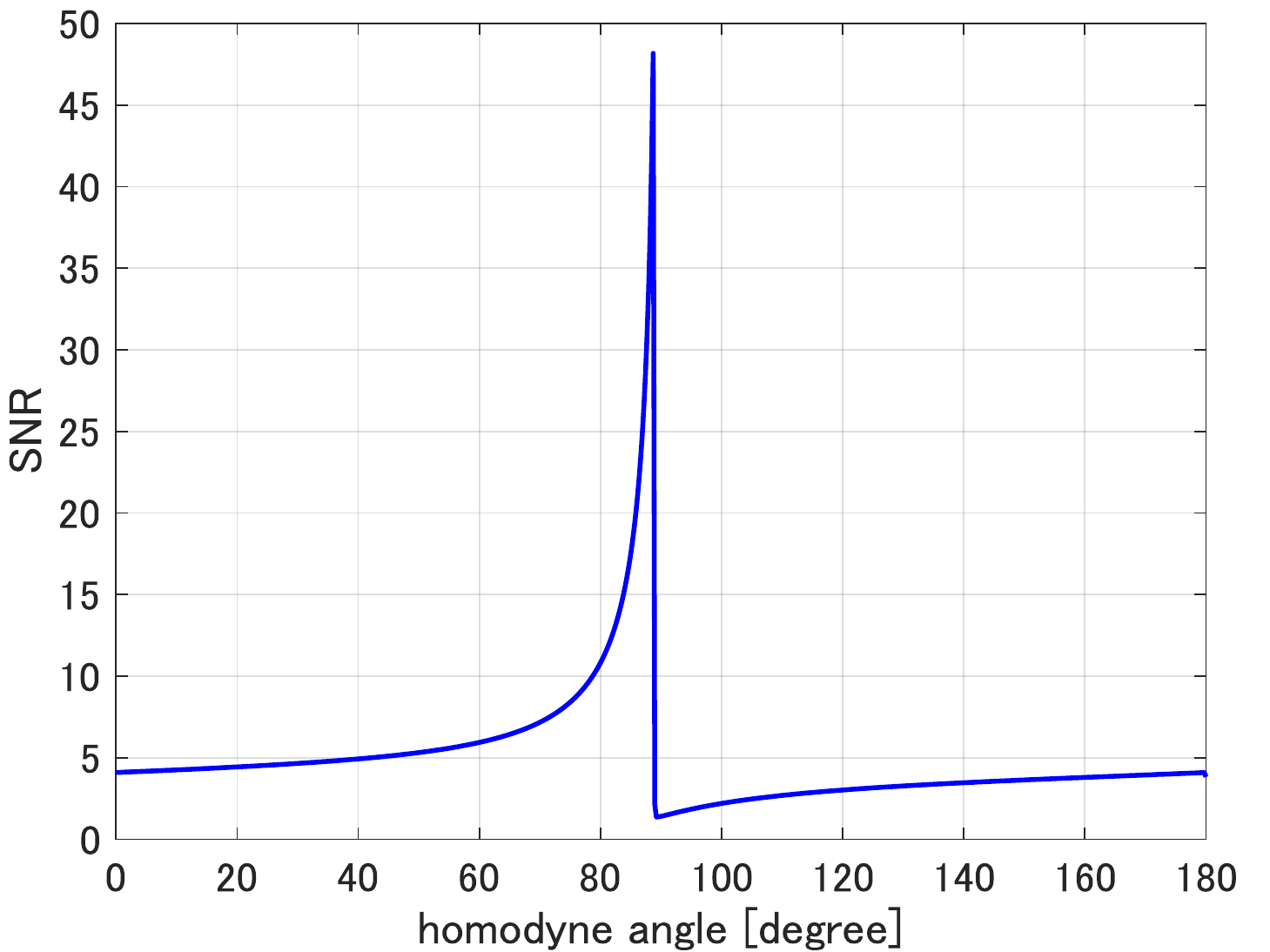}
 \subcaption{}
 \label{SNRa}
 \end{minipage}\\
\begin{minipage}{0.5\hsize}
 \includegraphics[width=35mm, bb=0 0 420 315]{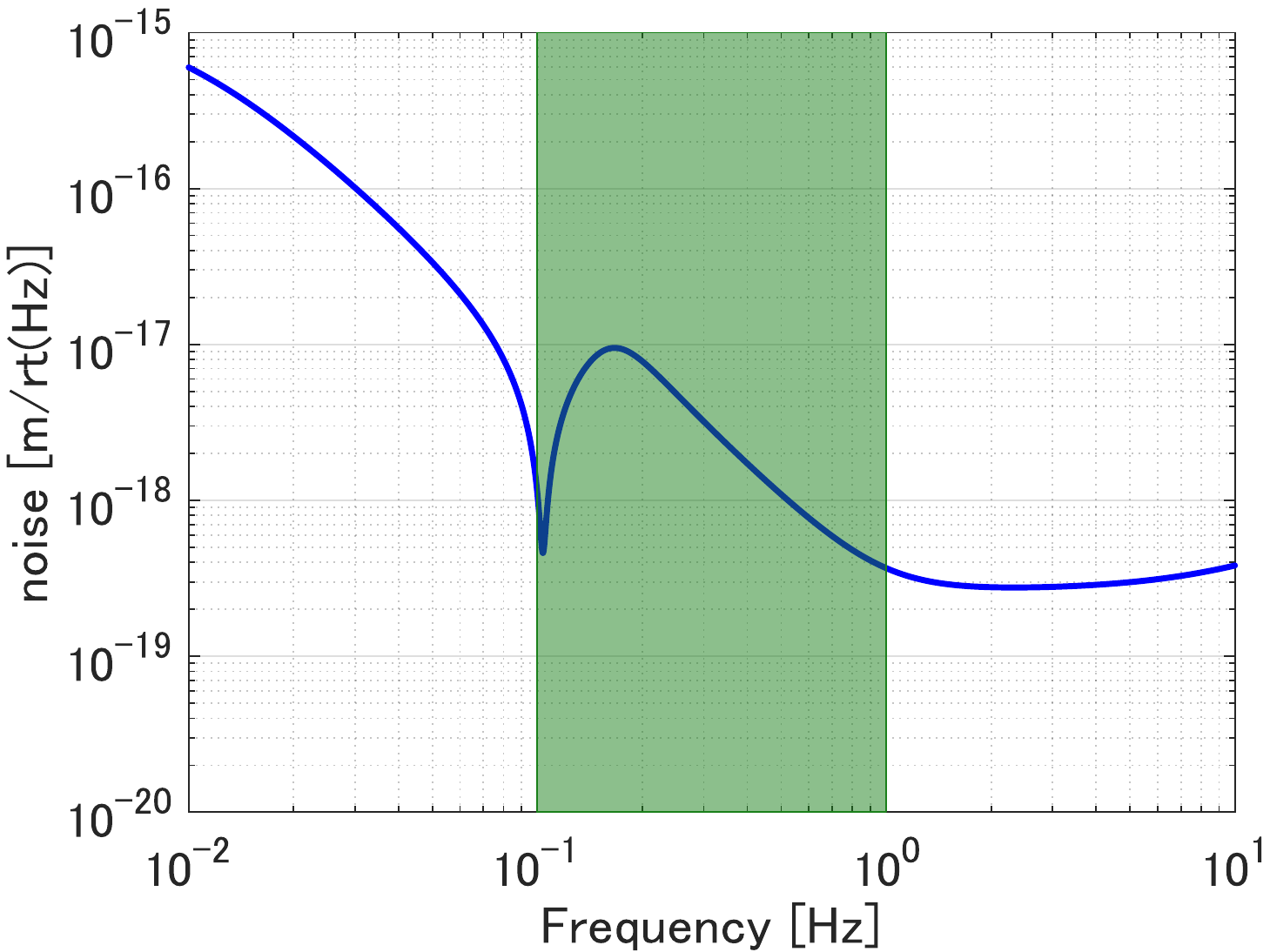}
 \subcaption{}
\label{SNRb}
\end{minipage}
\begin{minipage}{0.5\hsize}
 \centering
 \includegraphics[width=35mm, bb=0 0 420 315]{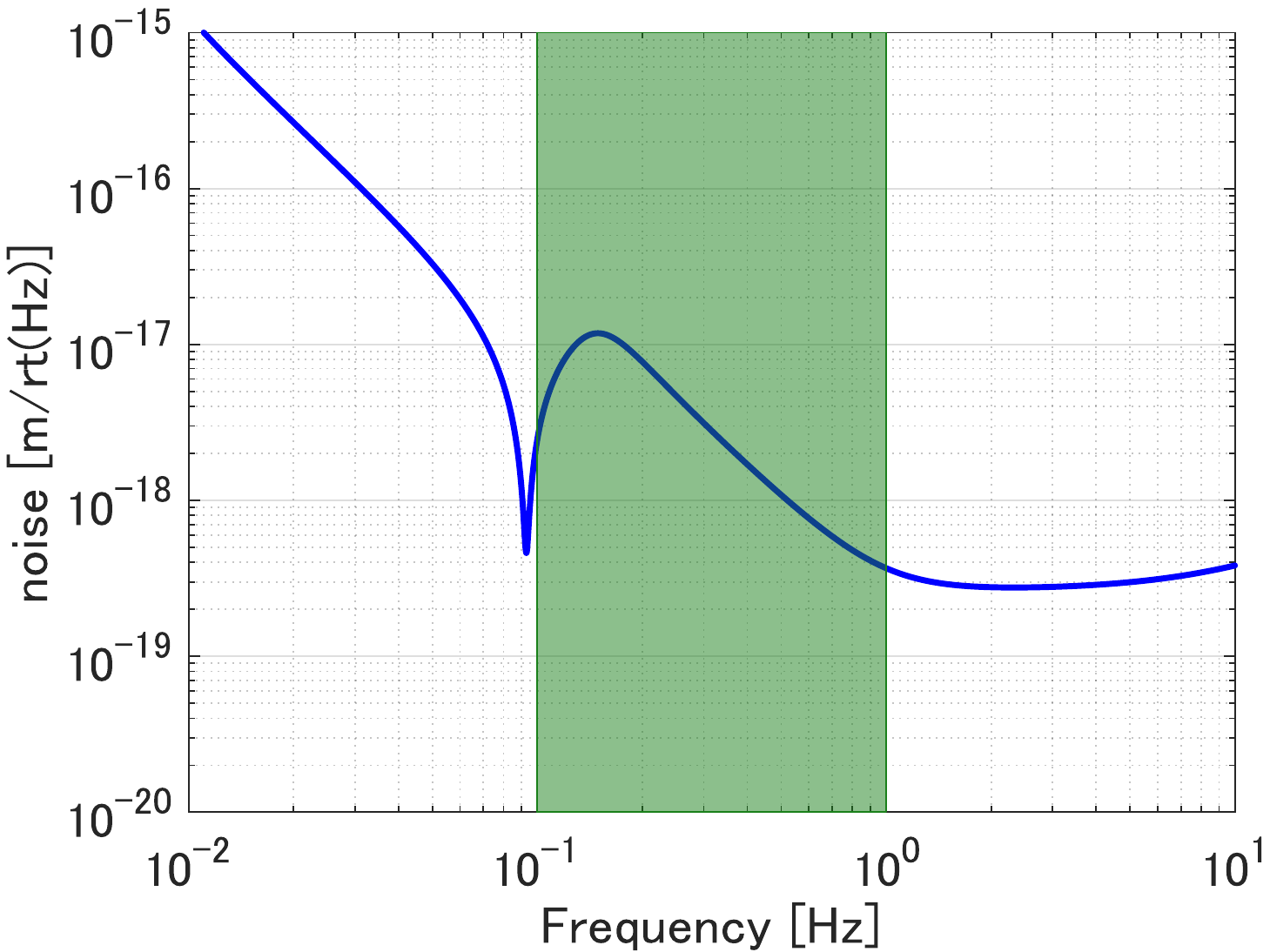}
 \subcaption{}
\label{SNRc}
\end{minipage}
\end{tabular}
\caption{(a) Homodyne angle dependence  on the SNR. (b) the total noise curve with a homodyne angle corresponding to the peak of the SNR in Panel~(a). (c) the total noise with a homodyne angle corresponding to the lowest SNR in Panel~(a).}
\label{SNR}
\end{figure}

Figure \ref{SNR} shows the SNR dependence on the homodyne angle. 
 With the completing-square method, the best SNR in the frequency band from 0.1 to 1 Hz is 33.2 with the best homodyne angle. This is 23.5 times better than the original noise that is obtained without quantum locking.
It should be noted that in the above estimate only the quantum noise is considered.
In reality,  some other noises are also expected to limit the sensitivity. Thus, the realistic best homodyne angle could be different. It should be also noted that the sharp peak in Fig.\ref{SNRa} indicates that the homodyne angle should be controlled with a high accuracy (better than 0.2 degree) to achieve high SNR (better than 40). We plan to experimentally demonstrate the feasibility of such an accurate control of the homodyne angle.

\section{Conclusions}

We proposed a method to optimize the quantum noise in presence of multiple interferometer outputs, independently of feedback gain, which is based on the "completing the square technique". Using this method, we optimized the quantum noise in a system where the quantum locking technique was used. Then we successfully found the best homodyne angle which optimized the total noise by maximizing the SNR with respect to the primordial gravitational waves on the basis of the optimized total noise. In this sense, this method was found to be highly effective. Furthermore, this method can be used in general cases to optimize the quantum noise when there are two or more outputs in the system.

\section*{Acknowledgments}
We would like to thank Masayuki Nakano for helpful discussion. We would like to thank Michael E. Zucker for the English editing. This work was supported by Daiko Foundation and JSPS KAKENHI Grant Number JP19H01924.





\end{document}